\documentstyle[epsfig,12pt]{article}
\newcommand{\be}{\begin{eqnarray}}
\newcommand{\ee}{\end{eqnarray}}

\def\lsim{\mathrel{\rlap{\lower4pt\hbox{\hskip1pt$\sim$}}
    \raise1pt\hbox{$<$}}}               
\def\gsim{\mathrel{\rlap{\lower4pt\hbox{\hskip1pt$\sim$}}
    \raise1pt\hbox{$>$}}}               

\begin{document}

\begin{figure}[htb]

\epsfxsize=6cm \epsfig{file=logo_INFN.epsf}

\end{figure}

\vspace{-4.75cm}

\Large{\rightline{Sezione ROMA III}}
\large{
\rightline{Via della Vasca Navale 84}
\rightline{I-00146 Roma, Italy}
}

\vspace{0.6cm}

\rightline{INFN-RM3 98/11}
\rightline{December 1998}

\normalsize{}

\vspace{1.5cm}

\begin{center}

\Large{Bloom-Gilman duality of the nucleon structure function\\[3mm] 
and the elastic peak contribution\footnote{\bf Published in Few-Body 
Systems Supplementum 10, 423-426 (1999), Proceedings of the XVI European 
Conference on {\em Few-Body Problems in Physics}, Autrans (France), 
July 1998.}}

\vspace{1.5cm}

\large{G. Ricco$^{(1)}$, M. Anghinolfi$^{(1)}$, M. Ripani$^{(a)}$, 
M. Taiuti$^{(1)}$ and S. Simula$^{(2)}$}

\vspace{1cm}

\normalsize{$^{(1)}$Physics Dept., University of Genova and INFN, 
Sezione di Genova, Italy\\[2mm] $^{(2)}$Istituto Nazionale di Fisica 
Nucleare, Sezione Roma III, Italy}

\end{center}

\vspace{1cm}

\begin{abstract}

\noindent The occurrence of the Bloom-Gilman duality in the nucleon structure
function is investigated by analyzing the $Q^2$-behavior of low-order
moments, both including and excluding the contribution arising from the
nucleon elastic peak. The Natchmann definition of the moments has been
adopted in order to cancel out target-mass effects. It is shown that the
onset of the Bloom-Gilman duality occurs around $Q^2 \sim 2 ~ (GeV/c)^2$ if
only the inelastic part of the nucleon structure function is considered,
whereas the inclusion of the nucleon elastic peak contribution leads to
remarkable violations of the Bloom-Gilman duality. 

\end{abstract}

\newpage

\rightline{}

\newpage

\setcounter{page}{1}

\pagestyle{plain}

\indent The investigation of inelastic lepton scattering off nucleon (and
nuclei) can provide relevant information on the concept of parton-hadron
duality, which deals with the relation among the physics in the
nucleon-resonance and Deep Inelastic Scattering ($DIS$) regions. As is well
known, well before the advent of $QCD$, local parton-hadron duality was
observed empirically by Bloom and Gilman \cite{BG70} in the proton structure
function $F_2(x, Q^2)$ measured at $SLAC$ (where $x \equiv Q^2 / 2m \nu$ is
the Bjorken scaling variable, $m$ the nucleon mass and $Q^2$ the squared
four-momentum transfer). More precisely, they found that the smooth scaling
curve measured in the $DIS$ region at high $Q^2$ represents a good average
over the resonance bumps seen in the same $x$ region at low $Q^2$.

\indent In Ref. \cite{duality} we addressed the specific question whether and
to what extent the Bloom-Gilman duality already observed in the proton
occurs also in the structure function of a nucleus. To that end all the
available experimental data for the structure functions of proton, deuteron
and light complex nuclei were analyzed in terms of low-order moments in the
$Q^2$ range from $0.3$ to $5 ~ (GeV/c)^2$. If only the inelastic parts of
the structure functions are considered, we found that in case of the proton
and the deuteron the Bloom-Gilman duality is fulfilled starting from $Q^2
\sim 2 ~ (GeV/c)^2$, whereas in case of complex nuclei, despite the poor
statistics of the available data, the onset of the local parton-hadron
duality is clearly anticipated. Besides these interesting findings, we
observed also that the inclusion of the contribution arising from the
nucleon elastic peak leads to remarkable violations of the local
parton-hadron duality for all the targets considered. In this contribution
we present an improvement of the work of Ref. \cite{duality} about the
failure of local parton-hadron duality around the nucleon elastic peak.

\indent Following the works of Refs. \cite{GP76,RGP77}, the analysis of Ref.
\cite{duality} was carried out using the Cornwall-Norton definition of the
moments, viz.
  \be
    \label{CN}
    M_n^{(CN)}(Q^2) \equiv \int_0^1 d\xi ~ \xi^{n - 2} ~ F_2(\xi, Q^2)
 \ee
where $\xi \equiv 2x / [1 + \sqrt{1 + 4 m^2 x^2 / Q^2}$ is the Nachtmann
variable. The $Q^2$ behavior of Eq. (\ref{CN}) was compared with the one of
the moments $A_n^{(CN)}(Q^2)$ of the leading-twist (dual) structure
function, viz.
 \be
    \label{CN-LT}
    A_n^{(CN)}(Q^2) \equiv \int_0^1 d\xi ~ \xi^{n - 2} ~ F_2^{(dual)}(\xi,
    Q^2)
\ee
with
 \be
    \label{dual}
    F_2^{(dual)}(\xi, Q^2) & = & {x^2 \over (1 + {4 m^2 
    x^2 \over Q^2})^{3/2}} {F_2^{(LT)}(\xi, Q^2) \over \xi^2} \nonumber \\
    & + & 6 {m^2 \over Q^2} {x^3 \over (1 + {4 m^2 x^2
    \over Q^2})^2} \int_{\xi}^1 d\xi' {F_2^{(LT)}(\xi', Q^2) \over \xi'^2}
    \nonumber \\
    & + & 12 {m^4 \over Q^4} {x^4 \over (1 + {4 m^2 x^2
    \over Q^2})^{5/2}} \int_{\xi}^1 d\xi' \int_{\xi'}^1 d\xi" 
    {F_2^{(LT)}(\xi", Q^2) \over \xi"^2}
 \ee
where the $\xi$-dependence as well as the various integrals appearing in
the r.h.s. account for target mass effects, which have to be included in
order to cover the low $Q^2$ region \cite{GP76}. In Eq. (\ref{dual})
$F_2^{(LT)}(x, Q^2)$ represents the leading-twist ($LT$) nucleon structure
function, fitted to high $Q^2$ proton and deuteron data, and extrapolated
down to low values of $Q^2$ by means of the Altarelli-Parisi evolution
equations. In the $DIS$ region one gets $F_2^{(LT)}(x, Q^2) = \sum_f e_f^2 x
\left [ \rho_f(x, Q^2) + \bar{\rho}_f(x, Q^2) \right ]$, with $\rho_f(x,
Q^2)$ being the quark distribution of flavor $f$. 

\indent However, Eqs. (\ref{CN}-\ref{dual}) suffer from a well known mismatch,
because $\xi(x = 1) \equiv \xi_{max} = 2 / (1 + \sqrt{1 + 4 m^2 / Q^2}) < 1$.
Therefore, while the evaluation of the integral of Eq. (\ref{CN}) stops at
the physical threshold $\xi_{max} < 1$, corresponding to the elastic
end-point $x = 1$, the r.h.s. of Eq. (\ref{CN-LT}) requires the values of the
dual function (\ref{dual}) in the unphysical region $\xi_{max} \le \xi \le1$.
Consequently, the comparison of the experimental $M_n^{(CN)}(Q^2)$, obtained
using all the available data sets and including the nucleon elastic peak
contribution, with the dual moments $A_n^{(CN)}(Q^2)$ should be handled with
care. Since the mismatch originates from target-mass corrections (i.e.,
from the fact that $m \neq 0$) we now adopt a different definition of the
moments, which avoids completely the mismatch problem, namely we consider the
Natchmann definition of moments \cite{NAT73}, given by
 \be
    \label{NAT}
    M_n(Q^2) \equiv \int_0^1 dx ~ {\xi^{n + 1} \over x^3} ~ F_2(x, Q^2) ~ {3
    + 3 (n + 1) r + n (n + 2) r^2 \over (n + 2) (n + 3)}
 \ee
where $r \equiv \sqrt{1 + 4m^2x^2 / Q^2}$ ($\xi = 2x / (1 + r)$). Since
the r.h.s. of Eq. (\ref{NAT}) projects out the contributions of spin-$n$
operators \cite{NAT73}, the $Q^2$ behavior of $M_n(Q^2)$ has to be compared
with the one of the $LT$ moments $A_n(Q^2)$, defined as
 \be
   \label{LT}
    A_n(Q^2) \equiv \int_0^1 dx ~ x^{n - 2} ~ F_2^{(LT)}(x, Q^2)
 \ee
where the $LT$ structure function $F_2^{(LT)}(x, Q^2)$ does not include any
target-mass corrections by definition. Now, the evaluation of Eq. (\ref{LT})
requires the values of the structure function $F_2^{(LT)}(x, Q^2)$ only in
the physical region $0 \le x \le 1$. 

\indent As pointed out in Ref. \cite{RGP77} and described in Ref.
\cite{duality}, the Bloom-Gilman duality should manifest itself in the
dominance of the $LT$ moments $A_n(Q^2)$ in the $Q^2$ behavior of low-order
moments $M_n(Q^2)$ starting from a value $Q^2 \simeq Q_0^2$ almost
independent of the order $n$ of the moment (for high values of $n$ the
moments of the structure functions $F_2(x, Q^2)$ and $F_2^{(LT)}(x, Q^2)$
should differ because of the rapidly varying behavior of the
nucleon-resonances peaks). In other words we expect that the low-order
residual moments $\Delta M_n(Q^2) \equiv M_n(Q^2) - A_n(Q^2)$ become a small
fraction of the $LT$ moments $A_n(Q^2)$ for $Q^2 \gsim Q_0^2$. This is
exactly what we see for $Q^2 \gsim Q_0^2 \sim 2 ~ (GeV/c)^2$ in Fig. 1(a),
where some low-order residual moments $\Delta M_n(Q^2)$ have been evaluated
including only the inelastic data. However, when the contribution from the
nucleon elastic peak is added, the dominance of $A_n(Q^2)$ starts from
values of $Q^2$ which strongly depend upon $n$ for $n > 2$. As for $n = 2$,
since the second moment $M_2(Q^2)$ corresponds to the area of the structure
function, the dominance of the $LT$ second moment $A_2(Q^2)$ correspond to
the {\em global} parton-hadron duality, which holds with and without the
inclusion of the nucleon elastic peak contribution (see Fig. 1). Our results
imply that the parton-hadron duality holds for local averages of the nucleon
structure function over the resonance bumps, but the elastic peak. Note that
the applicability of the concept of local parton-hadron duality in the region
around the nucleon elastic peak was found to be critical also in Refs.
\cite{BG70,RGP77}. Finally, let us point out that results of the same quality
as those shown in Fig. 1 hold as well in case of the deuteron structure
function.

\begin{figure}[htb]

\vspace{1cm}

\centering{\epsfxsize=15.85cm \epsfig{file=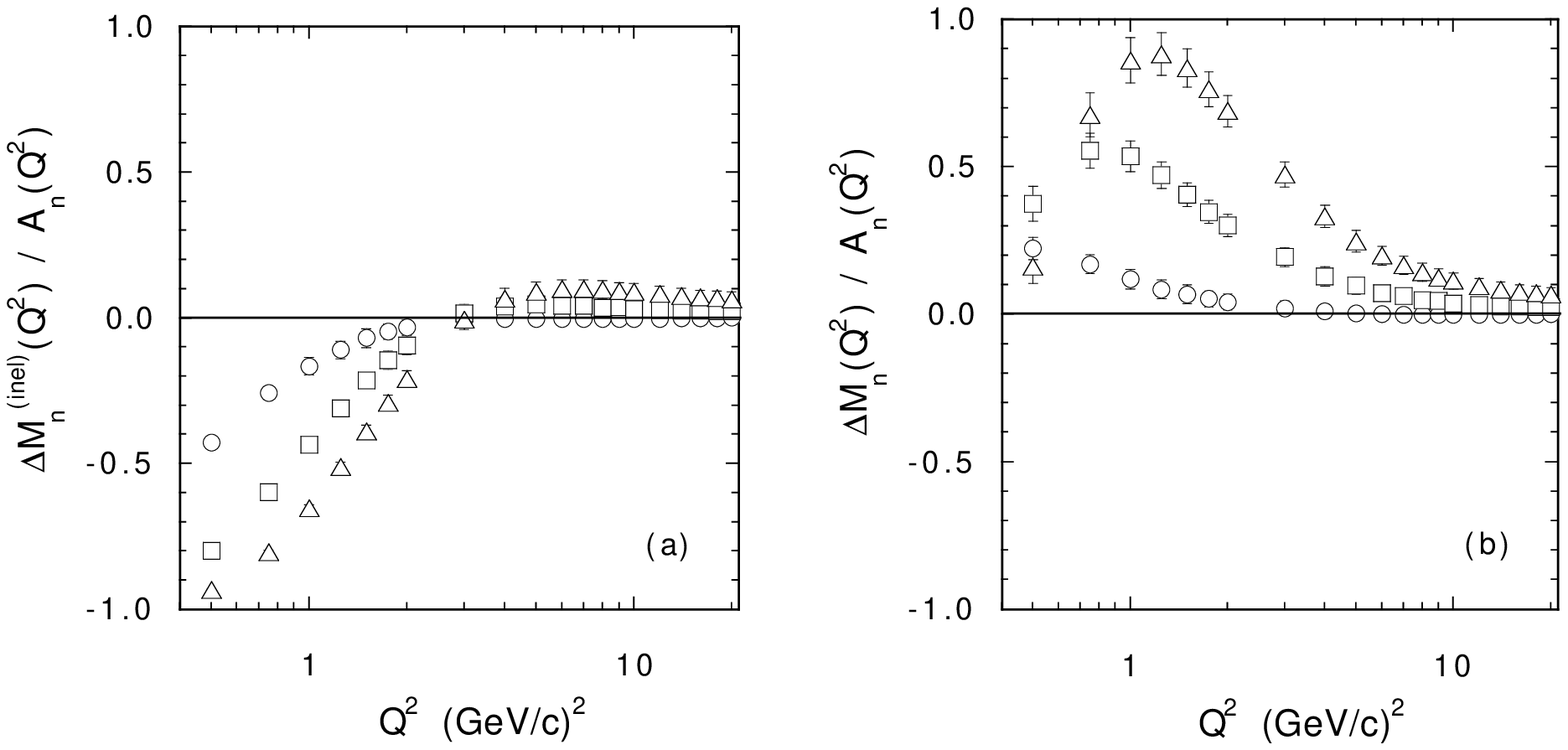}}

\vspace{0.2cm}

\parbox{0.05cm} \ $~~$ \ \parbox{15.70cm}{\small {\bf Figure 1}. Ratio of the
residual moments $\Delta M_n(Q^2) \equiv M_n(Q^2) - A_n(Q^2)$ to the
$LT$ moments $A_n(Q^2)$ (see Eqs. (\ref{NAT}-\ref{LT})) vs. $Q^2$. Dots,
squares and triangles correspond to $n = 2, 4$ and $6$, respectively. For
the calculation of $A_n(Q^2)$ (Eq. (\ref{LT})) the parton distributions of
Ref. \cite{GRV92} have been adopted. In (a) only the inelastic part of the
proton structure function is considered, whereas in (b) also the proton
elastic peak contribution is included in the determination of Eq.
(\ref{NAT}). For the experimental data set adopted for the evaluation of Eq.
(\ref{NAT}) see Ref. \cite{duality}.}

\vspace{1cm}

\end{figure}

\indent In conclusion, the occurrence of the Bloom-Gilman duality in the
nucleon structure function has been investigated by analyzing the
$Q^2$-behavior of low-order moments, both including and excluding the
contribution arising from the nucleon elastic peak. The Natchmann
definition of the moments has been adopted in order to cancel out
target-mass effects. It has been shown that the onset of the Bloom-Gilman
duality occurs around $Q^2 \sim 2 ~ (GeV/c)^2$ if only the inelastic part of
the nucleon structure function is considered, whereas the inclusion of the
nucleon elastic peak contribution leads to remarkable violations of the
Bloom-Gilman duality.

\end{document}